\documentstyle[epsfig,aps]{revtex}

\newcommand{\be}{\begin{eqnarray}}
\newcommand{\ee}{\end{eqnarray}}
\newcommand{\non}{\nonumber\\}
\newcommand{\bea}{\begin{eqnarray}}
\newcommand{\eea}{\end{eqnarray}}
\newcommand{\lton}{\mathrel{\lower.9ex
                  \hbox{$\stackrel{\displaystyle <}{\sim}$}}}

\begin{document}

\thispagestyle{empty}
\title {\bf Scattering of Gluons from the Color Glass Condensate}

\author
{
 Adrian Dumitru and Jamal Jalilian-Marian
 \\
 {\small\it Nuclear Theory Group, Physics Department, BNL, Upton, NY 11973}\\
}

\maketitle

\begin{abstract}

We prove that the inclusive single-gluon production cross section for
a hadron colliding with a high-density target factorizes into the gluon
distribution function of the projectile, defined as usual within the
DGLAP collinear approximation, times the cross section for scattering of
a single gluon on the strong classical color field of the target. We then
derive the gluon-proton (nucleus) inelastic cross section and show 
that it is (up to logarithms) infrared safe and that 
it grows slowly with center of mass energy.
Furthermore, we discuss jet transverse momentum broadening for the 
case of nuclear targets. We show that in the saturation regime, in contrast 
to the perturbative regime, the width of the transverse momentum 
distribution is infrared finite and grows rapidly with energy and rapidity. 
In both regimes, however, transverse momentum broadening exibits the same 
$A$ dependence.

\end{abstract}


\section{Introduction}

Understanding the behavior of hadronic cross sections at very high energy 
is one of the major unresolved problems in QCD. Even though Regge theory
can, in principle, predict the energy dependence of hadronic cross
sections and there are many phenomenological models, based on Regge
theory, which are somewhat successful in describing the data, the
relation of Regge theory to QCD is not well understood. Therefore, it 
would be very important to be able to calculate the high energy
behavior of hadronic cross section from QCD itself. However, it
is believed that total cross sections are intrinsically non-perturbative
and not amenable to perturbative QCD methods \cite{heb}. In this note, 
we consider
the simpler problem of (real) gluon-proton total inelastic cross section 
in the very high energy (small $x$) limit, using the effective action
and classical field method developed recently, and show that its growth
with energy is inhibited as compared with that expected from perturbative
QCD. 

At very small $x$, a hadron is a Color Glass Condensate due to
the condensation of gluons into a coherent state with characteristic
momentum of $Q_s(x)$
\cite{McLerran:1994ni,Mueller:1999wm,Kovchegov:1996ty,Jalilian-Marian:1997xn}.
In other words, most of the gluons in the wave 
function of a hadron have momenta of order $Q_s(x)$. As we go to higher
energies, $Q_s(x)$ increases and eventually will become much larger than
$\Lambda_{QCD}$ so that $\alpha_s(Q_s) \ll 1$ and weak coupling
methods can be used. Even though the theory may be weakly coupled, it
is still non-perturbative in the sense that one has high gluon
densities and strong classical color fields associated with them so
that the standard perturbative QCD breaks down. Much progress has been made in
developing a formalism which describes this
weakly coupled, though non-perturbative region of QCD. It generalizes the
standard collinear factorized (leading twist) expressions for hadronic 
cross sections and allows one to calculate hadronic cross sections
in an environment where higher twist (high gluon density at small $x$) 
effects are important. We refer the reader to \cite{larry} and
references therein for a review of this formalism. 

One can use this classical field method to calculate gluon production
at high energy~\cite{Kovner:1995ja}. In this Letter,
we use the results of \cite{dm} to prove (collinear)
factorization of the inclusive
cross-section for a ``dilute'' hadron impinging on a dense target.
Based on that result we then derive an expression for
the gluon-proton (nucleus) inelastic cross section and for
its energy dependence at very high energies. Using our results for
the gluon-proton (nucleus) inelastic cross section, we consider
the transverse momentum broadening of gluons due to
scattering from the strong classical field of a nucleus
and show that it is infared finite, energy dependent and scales like 
$A^{1/3}$.

\section{Gluon-proton inclusive cross section}

In the classical field and effective action approach to hadronic 
(nuclear) collisions at high energy, one solves the classical
Yang-Mills equations of motion in the presence of random color charges
created by the valence quarks and gluons at high $x$. One then averages
over these color charges with a Gaussian weight to compute physical
quantities. The classical fields of the colliding hadrons (nuclei)
before the collision are given by the single-hadron (nucleus) solutions
which, in light cone gauge, are
\bea
A^\pm_{1,2} &=& 0~,\non
A^i_{1,2} &=& \frac{i}{g} U_{1,2}(x_\perp)\partial^i U^\dagger_{1,2}(x_\perp)~.
\label{single}
\eea
These fields serve as the initial conditions for solving the Yang-Mills
equations of motion in the forward light cone; they were
solved in \cite{dm} for the case of asymmetric collisions where 
the classical field of one of the colliding sources is much stronger 
than the classical field of the other source.

\begin{figure}[htp]
\centerline{\hbox{\epsfig{figure=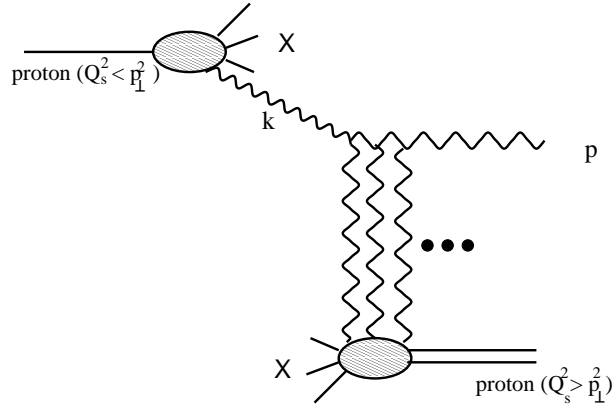,width=8cm}}}
\caption{Gluon production in inclusive $pp$ scattering at rapidity far
from the target proton (i.e., the strong color field).}
\label{fig_gprod}
\end{figure}
The produced gluon field in the forward light cone region  
is given by 
\bea
A^i(\tau,x_\perp) = U(x_\perp) \left( \beta^i(\tau,x_\perp) + \frac{i}{g}
\partial^i\right) U^\dagger(x_\perp)~,\non
A^\pm(\tau,x_\perp) = \pm x^\pm U(x_\perp) \beta(\tau,x_\perp)
 U^\dagger(x_\perp)~.  \label{sol_flc}
\eea
We have chosen the gauge condition $x^+ A^- + x^- A^+ =0$.
Here, $\tau=\sqrt{2x^+ x^-}$ denotes proper time and $x_\perp$ is the 
transverse coordinate.
The $U$'s are rotation matrices in color space, to be specified shortly.
At asymptotic times, $\tau\to\infty$, the fields $\beta$ and $\beta^i$ are
given by superpositions of plane wave solutions,
\bea
\beta(\tau\to\infty,x_\perp) &=& \int\frac{d^2p_\perp}{(2\pi)^2}
              \frac{1}{\sqrt{2\omega\tau^3}}
          \left\{
           a_1(p_\perp)e^{ip_\perp \cdot x_\perp-i\omega\tau} +c.c.\right\}~,\\
\beta^i(\tau\to\infty,x_\perp) &=& \int\frac{d^2p_\perp}{(2\pi)^2}
\frac{1}{\sqrt{2\omega\tau}}\frac{\epsilon^{il}p_\perp^l}{\omega}
          \left\{
          a_2(p_\perp)e^{ip_\perp \cdot x_\perp-i\omega\tau} +c.c.\right\}~.
\eea
The number distribution of produced gluons at rapidity $y$ and transverse
momentum $p_\perp$ is given by
\be \label{numdis}
\frac{dN}{d^2p_\perp dy} = \frac{2}{(2\pi)^2} {\rm tr}~\left(
\left| a_1(p_\perp)\right|^2 + \left| a_2(p_\perp)\right|^2\right)~.
\ee

One can obtain an analytical solution of the classical Yang-Mills 
equations if one of the fields (the ``projectile'')
is much weaker than the other (the ``target''). This situation is realized
physically when $|y-y_t|\gg|y-y_p|$, because of the renormalization group
evolution of the gluon density in
rapidity~\cite{Jalilian-Marian:1997xn,larry,yuri}.
In that case, it turns out~\cite{dm} that the
$U$'s appearing in eq.~(\ref{sol_flc}) are just the $U_2$'s from
eq.~(\ref{single}); that is, to leading order in the weak field, the plane
waves in the forward light cone are just gauge rotated by the strong field
\be
U_2(x_\perp,y) = {\cal P} \exp\left( -ig \int\limits_{y_t}^y dy'
\Phi_2(x_\perp,y')\right)~.
\ee
Here, we assumed that the target is moving along the negative $z$-axis, i.e.\
at negative rapidity $y_t<0$.
In the Color Glass Condensate model, we now have to
average the squared amplitudes over the gauge potentials
$\Phi_{1,2}(x_\perp,y)$ using a Gaussian
weight~\cite{McLerran:1994ni,Kovchegov:1996ty,Jalilian-Marian:1997xn}:
\bea \label{Gauss}
\langle |a_{1,2}|^2\rangle
&=& \int \!\! {\cal D} \Phi_1{\cal D} \Phi_2  \, |a_{1,2}(\Phi_1,\Phi_2)|^2
     \nonumber\\
&\times& \exp \left[
-\int\limits_y^{y_p} d y'\int d^2x_\perp \frac{{\rm tr} \left(\nabla^2_\perp
\Phi_1(x_\perp,y')\right)^2}{g^2\mu_1^2(x_\perp,y')}
- \int\limits_{y_t}^y d y'\int d^2x_\perp \frac{{\rm tr} \left(\nabla^2_\perp
\Phi_2(x_\perp,y')\right)^2}
{g^2\mu_2^2(x_\perp,y')}\right]~.
\eea
Here, $\mu$ denotes the density of color charge in the sources per unit of
transverse area and rapidity. The radiation number distribution~(\ref{numdis})
turns out to depend only on the {\em integrated} color charge densities of the
sources,
\be
\chi_1(y,p_\perp^2) = \int\limits_y^{y_p} dy' \mu^2_1(y',p_\perp^2)\quad,\quad
\chi_2(y,p_\perp^2) = \int\limits_{y_t}^y dy' \mu^2_2(y',p_\perp^2)~.
\ee
Due to the evolution in rapidity, $\chi_1(y)\ll\chi_2(y)$ if
$|y-y_t|\gg|y-y_p|$.
In terms of the gluon distribution function in the projectile or target
proton~\cite{Gyulassy:1997vt}, respectively,
\be \label{chi_def}
\chi_{1,2}(y,p_\perp^2) = 
\frac{N_c}{N_c^2-1} \frac{1}{\pi R^2} \int\limits^1_{x_0}
 dx \, g_{p,t}(x,p_\perp^2)~,
\ee
where $x_0$ is on the order of $\sim p_\perp\cosh(y)/\sqrt{s}$.
In the weak-projectile limit the averaging~(\ref{Gauss}) over the gauge
potentials with Gaussian weight can be done~\cite{dm}. At high transverse
momentum, $p_\perp\gg Q_t$ (where $Q_t$ is the saturation momentum of the
target), one recovers the standard result from perturbation
theory~\cite{Kovner:1995ja,Gyulassy:1997vt,KovRi},
\be
\frac{dN}{d^2p_\perp dy} = \frac{4\alpha_s^2}{\pi R^2}
 \frac{N_c^2}{N_c^2-1} 
\frac{\alpha_s N_c}{p_\perp^4} 
\int \frac{d^2 k_\perp}{\pi^2} \frac{p_\perp^2}{k_\perp^2
(p_\perp-k_\perp)^2}
\int dx' g_p(x',k_\perp^2) \int dx'' 
g_t(x'',(p_\perp-k_\perp)^2)~. \label{pert_distr}
\ee
This is to be expected, of course, because the gluon occupation numbers in the
target proton become small at $p_\perp\gg Q_t$, i.e.\ that field
becomes weak as well and so can be treated perturbatively.

In the collinear limit ($k_\perp/p_\perp \rightarrow 0$), the first factor 
in~(\ref{pert_distr}) is the DGLAP~\cite{DGLAP} splitting function for
gluons.
It evolves the gluon distribution function of the projectile
from the scale $k_\perp^2$ to the scale $p_\perp^2$,
\be \label{proj_DGLAP}
xg_p(x,p_\perp^2) = \alpha_s N_c 
\int\limits^{p_\perp} \frac{d^2 k_\perp}{\pi^2} 
\frac{1}{k_\perp^2}\int dx' g_p(x',k_\perp^2)~.
\ee
Thus, in the collinear limit eq.~(\ref{pert_distr}) simply turns into
\be
\frac{dN}{d^2p_\perp dy} = \frac{4\alpha_s^2}{\pi R^2}
 \frac{N_c^2}{N_c^2-1} \frac{1}{p_\perp^4}
 xg_p(x,p_\perp^2) \int dx'' 
g_t(x'',p_\perp^2)~. \label{pert_distr2}
\ee
Integrating over
the impact parameter space (simply a factor of $\pi R^2$) 
and dividing by the flux of incoming
gluons in the projectile proton at the hard scale $p_\perp^2$, which is simply
$xg_p(x,p_\perp^2)$, we obtain the inclusive gluon-proton differential
cross section at {\em large} momentum transfer
\be \label{pertXsection}
\frac{d\sigma^{\rm pert}_{gp}}{d^2p_\perp} =
4\alpha_s^2 \frac{N_c^2}{N_c^2-1} \frac{1}{p_\perp^4} \int\limits_{x_0}^1 dx \,
g_t(x,p_\perp^2)~.
\ee
This is the standard expression for the gluon-proton inclusive cross
section~\cite{Gyulassy:1997vt,BDMPS} in the one-gluon exchange approximation
which diverges like $1/t^2\equiv 1/p_\perp^4$ at small momentum transfer. 
$\sigma^{\rm pert}_{gp}$ grows like a power of energy if $xg(x)\sim
1/x^\delta$ with $\delta>0$. 

Next, we consider the radiation distribution~(\ref{numdis}) at small
transverse momentum, $p_\perp\lton Q_t$, but large enough so that the
projectile proton is still in the weak-field regime 
$p_\perp\gg \Lambda_{QCD}$. In this regime~\cite{dm}
\bea
\frac{dN}{d^2p_\perp dy} &=& \frac{1}{4\pi} \frac{1}{p_\perp^2}
\alpha_s N_c 
\int \frac{d^2 k_\perp}{\pi^2} \frac{p_\perp^2}
{k_\perp^2 (p_\perp-k_\perp)^2}
  \int\limits_{x_0}^1 dx\, g_p(x,k_\perp^2)\\
&\simeq& 
\frac{1}{4\pi} \frac{1}{p_\perp^2} \,
xg_p(x,p_\perp^2)~. \label{sat_distr}
\eea
This result is valid to leading order in $\alpha_s^2\chi_1$, but to all
orders in $\alpha_s^2\chi_2$. 
The second line applies in the limit of nearly collinear splitting
($k_\perp/p_\perp\to0$), where we can employ eq.~(\ref{proj_DGLAP}).
Thus, the inclusive gluon production again factorizes into the gluon
distribution function of the projectile at the scale $p_\perp^2$ times
the cross section for scattering of a gluon on the high-density target.

To obtain the gluon-proton cross section,
we again integrate over
the impact parameter and divide by the flux of incoming collinear gluons 
in the projectile at the scale $p_\perp^2$, which is
$xg_p(x,p_\perp^2)$. We get
\be \label{sat_diffXsection}
\frac{d\sigma^{\rm sat}_{gp}}{d^2p_\perp} =
\frac{1}{4\pi} \frac{1}{p_\perp^2} \pi R^2
\ee
In the saturation regime, the cross section is of order 1 rather than 
$\alpha_s^2$ as in~(\ref{pertXsection}) because the occupation number of the
target $\sim 1/\alpha_s$ cancels one power of the coupling in both the
amplitude and the complex conjugate amplitude.
Note that the above result holds for $p_\perp\gg \Lambda_{QCD}$ only.
To get the gluon-proton total inclusive cross section, we integrate
(\ref{sat_diffXsection}) which gives 
\be \label{sat_Xsection}
\sigma^{\rm sat}_{gp}
 \simeq \frac{1}{4} \pi R^2 \log \left(Q_t^2/\Lambda_{QCD}^2\right)
+{\cal O}(\alpha_s^2)~.
\ee

Parameterization of the saturation momentum like a power
of energy, $Q_t^2\sim s^\gamma$ as done in \cite{gbw} then 
leads to a logarithmically growing cross section
\be \label{areagrowth}
\sigma_{gp}
 \sim \pi R^2 \log s~.
\ee
On the other hand, if we use a DLA DGLAP type parameterization
$Q_t\sim \exp{\sqrt{\log 1/x}}$, the cross section would grow like
a square root of energy. Therefore, it is clear that different
parameterization of the target saturation scale $Q_t$ would lead to
a different growth of the cross section with energy. Curiously enough,
assuming $Q_t$ to grow like $s^{\log s}$ would lead to growth of the
gluon-proton cross section like $\log ^2 s$. In order to determine
the energy dependence of $Q_t$ rigorously, one would need to
formally define it in terms of a gluonic two-point function and solve
the non-linear evolution equation that $Q_t$ follows. This is however
beyond the scope of this work. 

Strictly speaking, our results are correct for the cross
section per unit area, ${d\sigma / d^2b}$. What we have shown here is
that the growth of this cross section at a fixed impact parameter is inhibited
due to high gluon density effects. In principle, as we go to larger
impact parameters $b$, the gluon density becomes smaller and smaller
until our classical formalism reduces to the standard pQCD where the
gluon density is not that large. Since large impact parameters (where
the gluon density is low and so the classical approach is not valid) are 
believed to give the dominant contribution to the total cross section,
one should understand the area appearing in~(\ref{areagrowth}) as a
parameter which we can not precisely determine. A genuine non-perturbative
(strong coupling vs.\ our weak coupling methods) calculation would presumably
enable one to determine this factor and its energy dependence.

The above equations hold equally well for a nuclear target. In
eq.~(\ref{pertXsection}) one just has to replace the gluon distribution
function of the target proton by that for the nucleus; in the absence of
shadowing~\cite{Mueller:1999wm,Frankfurt:nt}, $g_A(x,Q^2) = A g(x,Q^2)$.
On the other hand,
in eqs.~(\ref{sat_diffXsection},\ref{sat_Xsection}) one substitutes the
radius of the proton by that of the nucleus, $R_A\simeq A^{1/3}R$.

It is now straightforward to compute the transverse momentum broadening 
of the incoming gluon jet traversing the target nucleus~\cite{BDMPS,Luo:fz}.
The transverse momentum broadening is given by~\cite{BDMPS}
\be \label{def_broad}
\left<p^2_{\perp}\right> = \frac{1}{A}\left< t_A(\vec b)\right>
\int d^2p_\perp \, p^2_\perp \frac{d\sigma_{gA}}{d^2p_\perp}~,
\ee
where $\left<t_A(\vec b)\right>$ is the nuclear thickness function averaged
over impact parameters:
\be \label{def_tA}
\left< t_A(\vec b)\right> = \frac{A}{\pi R_A^2} = \rho L~,
\ee
with $L$ the average thickness of the nucleus,
and with $\rho\simeq 0.15$~fm$^{-3}$ the density of nucleons in the nucleus.

In perturbation theory, using~(\ref{pertXsection}) and DGLAP~\cite{DGLAP}
evolution,
\be
\frac{\alpha_s N_c}{\pi} \int dx\,g_A(x,p_\perp^2) =
\frac{d}{d\log p_\perp^2} xg_A(x,p_\perp^2)~,
\ee
one obtains
\be \label{pert_broad}
\left<p^2_{\perp}\right> = 4\pi^2\,\alpha_s\,\frac{N_c}{N_c^2-1} 
\left[xg(x,Q^2_{max})-xg(x,Q^2_{min})\right] \rho L~.
\ee
Thus, $\left<p^2_{\perp}\right>$ grows proportional to $A^{1/3}$ if 
nuclear shadowing is disregarded~\cite{BDMPS,Luo:fz}. Appearance of
$xg(x,Q^2_{min})$ signifies sensitivity to physics at small momentum transfer.
For a discussion of what value of $x$ is to be used 
in~(\ref{pert_broad}), see~\cite{BDMPS,Luo:fz}.

Transverse momentum broadening in the saturation regime is quite different.
Using~(\ref{sat_diffXsection}) and~(\ref{def_broad},\ref{def_tA}), we 
find that
\be
\left<p^2_{\perp}\right> \simeq \frac{Q_A^2}{4}+{\cal O}(\alpha_s)~,
\ee
where $Q_A^2$ denotes the nuclear saturation scale which, from the 
definition of $\chi_2\sim Q_A^2/\alpha_s^2$ in~(\ref{chi_def}), is 
$A^{1/3}$ times larger than that for a proton. Thus, in the saturation 
regime jet broadening grows with the {\em same} power of the nuclear 
mass number $A$ as in the perturbative regime 
$\left<p^2_{\perp}\right>\sim A^{1/3}$. Also, our derivation shows that
the jet transverse momentum broadening is energy dependent even though
we can not determine its energy dependence in a model independent way.
However, for $Q_A^2\sim s^\gamma$~\cite{gbw}, we obtain a strong power-law
increase of transverse momentum broadening. We would like to emphasize
\cite{djm} that since the saturation scale $Q_t^2$ of the target
is expected to be much larger in
the forward rapidity (projectile fragmentation) region, the jet transverse
momentum broadening will be much larger in the forward region than in
the central rapidity region, contrary to the prediction~(\ref{pert_broad})
from perturbation theory.

\section{Conclusion}
In summary, we have shown that inclusive gluon production from a hadron
scattering off a high-density target (with ``saturated'' gluon density)
factorizes into the gluon distribution of the projectile
times the cross section of the beam of incoming collinear gluons
on the dense target.
We derived the gluon-proton inclusive cross section
in the high energy limit. We have shown that the cross section grows,
with reasonable, HERA compatible parameterizations \cite{gbw} of the 
saturation momentum,  
only logarithmically with energy rather than power like as expected from 
perturbation theory. This ``unitarization'' of the cross section is due to 
the strong classical fields of the target generated by the high gluon 
density (higher twist) effects. We have also considered the transverse 
momentum broadening of the gluon jet passing through a nuclear target.
We have shown that it scales like $A^{1/3}$ in both perturbative and 
saturation regimes and that it is infrared finite. We predict that
the jet transverse momentum broadening will be larger in the forward rapidity
region and that it increases with energy.

\section*{Acknowledgement}
We thank Yu.\ Dokshitzer, D.\ Kharzeev, and Yu.\ Kovchegov for helpful
comments and L.\ McLerran for many stimulating discussions and a critical
reading of this manuscript. This manuscript has been authored under contract 
No.\ DE-AC02-98CH10886 with the U.S.\ Department of Energy.
J.J-M.\ is also supported in part by a LDRD from BSA.

\end{document}